\documentclass[12pt]{article}
\usepackage{epsfig}

\newcommand{\be}{\begin{eqnarray}}
\newcommand{\ee}{\end{eqnarray}}
\newcommand{\sll}{\raise.15ex\hbox{$/$}\kern-.43em\hbox{$l$}}
\newcommand{\slp}{\raise.15ex\hbox{$/$}\kern-.43em\hbox{$p$}}
\newcommand{\slq}{\raise.15ex\hbox{$/$}\kern-.43em\hbox{$q$}}
\newcommand{\slk}{\raise.15ex\hbox{$/$}\kern-.43em\hbox{$k$}}
\newcommand{\slepsilon}{\raise.15ex\hbox{$/$}\kern-.53em\hbox{$\epsilon$}}

\newcommand{\lsim}{\mbox{\raisebox{-0.6ex}{$\stackrel{<}{\sim}$}}\:}

\begin{document}

\bibliographystyle{unsrt}
\footskip 1.0cm

\thispagestyle{empty}
\begin{flushright}
INT--PUB 05--14
\end{flushright}
\vspace{0.1in}

\begin{center}{\Large \bf {The Color Glass Condensate and hadron
      production in the forward region}}\\

\vspace{1in}
{\large  Adrian Dumitru$^a$, Arata Hayashigaki$^a$ and 
  Jamal Jalilian-Marian$^b$}\\

\vspace{.2in}
{\it 
$^a$Institut f\"ur Theoretische Physik, J.~W.~Goethe Universit\"at,\\
Max-von-Laue Strasse 1,\\
D-60438 Frankfurt am Main, Germany\\
$^b$Institute for Nuclear Theory, University of Washington,
Seattle, WA 98195 }

\end{center}

\vspace*{25mm}

\begin{abstract}

\noindent 
We consider one loop corrections to single inclusive particle
production in parton-nucleus scattering at high energies, treating the
target nucleus as a Color Glass Condensate. We prove by explicit
computation that in the leading $\log Q^2$ approximation, these
corrections lead to collinear factorization and DGLAP evolution of the
projectile parton distribution and hadron fragmentation functions.
In single-inclusive cross sections,
only two-point functions of Wilson lines in the adjoint and
fundamental representations (Mueller's dipoles) arise, which can be
obtained from the solution of the JIMWLK equations.  The application
of our results to forward-rapidity production shows
that, in general, recoil effects are large. Hence, the forward
rapidity region at RHIC is rather different from the central region at
LHC, despite comparable gluon densities in the target. We show that
both the quantum $x$-evolution of the high-density target as well as
the DGLAP $Q^2$-evolution of the parton distribution and fragmentation
functions are clearly seen in the BRAHMS data. This provides additional
strong evidence for the Color Glass Condensate at RHIC.

\end{abstract}
\newpage

\section{Introduction}

The rapidity dependence of the recent RHIC data on hadron production
in deuteron-gold collisions~\cite{rhic} may hint at the emergence of the Color
Glass Condensate (CGC)~\cite{nonlin}
as the dominant physics in the forward
rapidity region~\cite{adjjm_prl}. While the ratio of the hadron
transverse momentum distributions
in $d+Au$ versus $pp$ collisions shows a Cronin enhancement at
midrapidity and
moderate transverse momentum~\cite{multi}, the enhancement turns into
suppression as one goes to larger rapidities.  The disappearance of
the Cronin peak is most commonly taken to be due to the quantum
evolution (with $\log 1/x$) of gluons in the target~\cite{forsup,dima,jjm}. 
Another argument in favor of the CGC
formalism is provided by the changing centrality
dependence of the data as one goes to the forward region.
Note also that leading-twist next-to-leading order perturbative QCD
calculations~\cite{WernerMark} provide a good description of the
inclusive distribution at forward rapidities in $pp$ collisions but
fail to describe the $d+Au$ BRAHMS $h^-$ data by a conventional modification
of the leading-twist parton densities in nuclei (shadowing).
The CGC dynamics can be tested further by $p+A$ collisions at the LHC,
where measurements in both the central and forward rapidity regions
should be performed. Interestingly, particle
production in the forward region of hadron-nucleus collisions
also plays a key role for the properties of giant air showers from
ultra-high energy cosmic ray interactions in the
atmosphere~\cite{DDS}. Improving our understanding of particle
production at large Feynman-$x$ may therefore reveal
the type and origin of those super-high energy particles.

Even though the qualitative predictions of the CGC were confirmed by the
RHIC data, in order to firmly establish the CGC as the cause
for the observed suppression of the hadron spectra in
deuteron-gold collisions and to clarify the role of other
scenarios~\cite{rudi} without $\log 1/x$ resummation, one needs to
consider effects in the CGC framework which have been
neglected so far and which may be significant. One such effect is the recoil
of the source which radiates the produced gluons. Current approaches
to gluon (hadron) production in proton-nucleus
collisions~\cite{CGCg_norecoil1,CGCg_norecoil2,CGCg_norecoil3}
treat both the projectile proton and the target nucleus as a Color Glass
Condensate, with the difference that the proton is assumed to be in
the dilute regime while the target nucleus could be in either the
dilute or dense regimes. This approach neglects recoil effects since
only diagrams that survive in the $\xi\to0$
limit, with $\xi$ the momentum fraction of the produced gluon, are included.
Gluon radiation and recoil effects should also be significant for the
production of leading hadrons in deep-inelastic scattering from
nuclei, where large distortions of the spectrum relative to
leading-twist calculations have been predicted~\cite{FGMS,Bartels:2003aa}.

However, in forward rapidity production of particles, the momentum
fraction $x$ of the projectile parton is large and treating the proton
as a dilute CGC (i.e.\ subject to BFKL evolution~\cite{bfkl}) can not be
justified. Theoretically, BFKL resummation applies if the
parameter $\alpha_s \, \log x_0/x \sim 1$, where $x_0 \sim 0.1$ is
where the valence degrees of freedom reside. In order to gain
appreciable longitudinal phase space so that BFKL evolution becomes
significant, one needs to go much below $x_0$, which is not the case
in the forward rapidity region, where (by definition) the typical
$x$ in the proton wave function is of order $x_0$. It is
known from HERA data that the saturation scale of a proton
reaches $\sim1$~GeV only around $x \sim 10^{-4}$.  Since
the saturation scale of the proton is so small (of order of
$\Lambda_{QCD}$) at $x \sim x_0$, even the extended scaling regime is
rather small, as well. The production of hadrons with
$p_t\gg\Lambda_{QCD}$ in
the forward rapidity region never receives any large
contributions from either the saturation or the extended scaling
regimes of the proton.

We therefore treat the projectile proton as a collection of quarks and
gluons according to the parton model and consider scattering of these
partons from the target nucleus, which is
treated as a Color Glass Condensate.  In Section~2 we
consider one loop corrections to quark-nucleus scattering due to
gluon radiation and show that recoil effects can be large. We then
generalize this to include all processes at this order and show that
in the leading $\log Q^2$ approximation, these lead to DGLAP
evolution~\cite{dglap}
of the quark and gluon distribution functions of the projectile
proton, and of their fragmentation functions into hadrons. 
Thus, we prove explicitly that one can treat a high energy
proton-nucleus collision as scattering of collinearly factorized
partons (which evolve according to DGLAP evolution equation) in the
proton on a dense nucleus treated as a Color Glass Condensate.

Furthermore, we show that the standard genuinely non-Abelian diagram
where the produced gluon scatters from the saturated target field
gives a contribution which decreases by a factor of two toward large
rapidity. We then show that QED-like bremsstrahlung, which is usually
disregarded, contributes about equally at large rapidity and ``small''
transverse momentum. Both contributions involve only two-point
functions of Wilson lines
(dipoles~\cite{Mueller:1994jq,Bartels:2003aa}), in the
adjoint and fundamental representations, respectively.
This makes it possible to
use the RHIC data on hadron production in $d+Au$ collisions to make
predictions for electromagnetic processes such as photon and dilepton
production~\cite{multi,Baier1} since these electromagnetic processes also
involve dipoles in the fundamental representation.  The evolution of
the target wave function with $x$ can then be included by using the
solution of the JIMWLK equations~\cite{nonlin} for the dipole
evolution (for first attempts see e.g.~\cite{Heribert}) or
alternatively, by using phenomenological parameterizations of the
dipole profile.

Finally, we apply our results to forward rapidity hadron production in 
deuteron-gold collisions at RHIC energy. Lacking a solution of the
JIMWLK equations we resort to the phenomenological parametrization of
Kharzeev, Kovchegov and Tuchin for the dipole profile~\cite{dima}. We find that
the BRAHMS data can be reproduced with a $p_t$-{\em independent}
$K$-factor. We show that both the DGLAP $Q^2$-evolution of the
distribution and fragmentation functions as well as the quantum
$x$-evolution of the high-density target (anomalous dimension of its gluon
distribution function) are clearly seen in the data. In our opinion,
this strengthens the evidence for the Color Glass
Condensate in high-energy $d+Au$ collisions~\cite{KLN} substantially.

Despite the rather good description of the data provided by the
above-mentioned formalism, some words of caution may be in order. It
has been shown within the NLO pQCD framework that at
RHIC energy, incoherent leading-twist interactions with partons from
the nucleus about $x_A\sim0.01$ contribute
significantly~\cite{WernerMark} (in fact, even overshooting the data at
forward rapidity). While the contribution from this region can be
neglected when very small $x$ in the nucleus become accessible at very
high energies, at RHIC energy it might be necessary to think about
additional mechanisms which suppress such interactions~\cite{FSprivcom}.
In this paper we do not construct any such model, however, but assume
the validity of the high-energy approximation inherent in the
non-linear $x$-evolution, namely that the target fields at rapidities
below that of the probe can be integrated out to yield the effective
saturation scale $Q_s(y)$. We return to this point in
appendix~\ref{appendixB}, where we show that the $2\to1$ like
kinematics which arises in the high-energy limit probes very small
momentum fractions in the target nucleus, on the order of $\langle
x_A\rangle \sim 10^{-3}$ for hadrons produced at rapidity $Y\simeq3.2$
(BRAHMS) and $\langle x_A\rangle \sim 10^{-4}$ for $Y\simeq4$ (STAR).

\section{Including recoil in gluon production}

\begin{figure}[htp]
\centering
\setlength{\epsfxsize=8cm}
\centerline{\epsffile{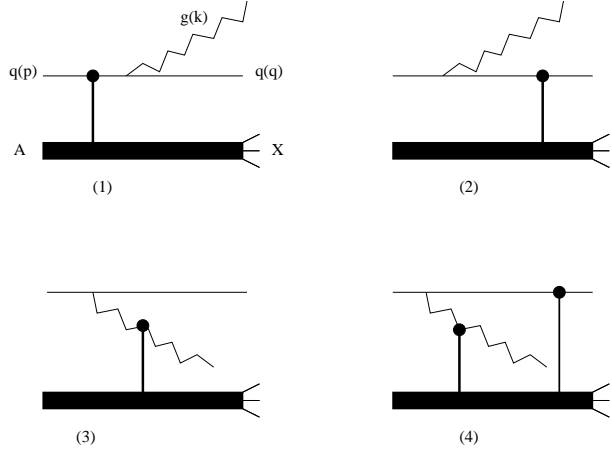}}
\caption{Diagrams contributing to gluon production in quark-nucleus
scattering.}
\label{fig:diagrams}
\end{figure}
The diagrams contributing to gluon production in quark-nucleus
scattering are shown in Fig.~\ref{fig:diagrams}. The
amplitudes for production of a massless quark with momentum $q$ and a
gluon with momentum $k$ are given in~\cite{jjmyk} and read
\be
 M_1 &=&- i g {1 \over 2q\cdot k}  \bar{u}(q)\, \slepsilon\, 
(\slq + \slk)\,\gamma^-\,u(p)\,
 t^a\,[V(q_t + k_t) - (2\pi)^2 \delta^2(q_t + k_t)]\nonumber\\
M_2 &=& i g {1 \over 2p\cdot k}  \bar{u}(q)\, \gamma^- \, 
(\slp - \slk)\,\slepsilon \,u(p)\,
 [V(q_t + k_t) - (2\pi)^2 \delta^2(q_t + k_t)]\,t^a\nonumber\\
M_3 &=& i g {k^- \over p\cdot q}  \bar{u}(q)\, \gamma_{\nu}\,u(p)\, 
d^{\nu\mu} (p-q)\, \epsilon_\mu (k) \,t^b\,[
U^{ba}(q_t+k_t) - \delta^{ba}\,(2\pi)^2 \delta^2(q_t + k_t)]\nonumber\\ 
M_4 &=& i g {k^- \over p^-} \int  {d^2l_t \over (2\pi)^2} \bar{u}(q) 
\,\gamma^-\,(\slp - \sll)\,\gamma_{\nu}\,u(p)\,
{d^{\nu\mu} (l) \over l_t^2} \, \epsilon_\mu (k) \nonumber \\
&& [V(q_t + l_t) - (2\pi)^2 \delta^2(q_t + l_t)]
t^b [U^{ba}(k_t - l_t) - \delta^{ba}\,(2\pi)^2 \delta^2(k_t - l_t)]~,
\label{eq:amp_1234_mom}
\ee
where $p (q)$ is the momentum of the incoming (outgoing) quark and $z\equiv
q^-/p^-$ the fractional light-cone momentum carried by the quark in
the final state ($\xi\equiv 1-z=k^-/p^-$ is that of the gluon). All other
possible diagrams are suppressed by powers of energy and will be ignored.
\be
V(x_t) &\equiv& \hat{P} \,\exp \left({ig\int dz^-\, A^+_a (x_t,z^-)
  \,t_a} \right)~,
\label{vdef} \\ 
U (x_t) &\equiv& \hat{P} \, \exp \left({ig\int dz^-\, A^+_a (x_t,z^-)
  \, T_a} \right)
\label{udef} 
\ee
are Wilson lines, in the fundamental and adjoint representations,
respectively, running along the light cone and summing up the
non-Abelian phases of the colored particles propagating through the
color-field of the nucleus.
To calculate the gluon production cross section, we square
the sum of the amplitudes~(\ref{eq:amp_1234_mom}) and then integrate over the
momentum of the outgoing quark, $q_t$.

After squaring the amplitude in eq.~(\ref{eq:amp_1234_mom}), it can be 
shown~\cite{aaj} that some of the terms exhibit
a collinear singularity upon integration over the transverse
momentum $q_t$ of the final state quark, which arises when the quark
and gluon in the final state are collinear\footnote{Care should be
  taken not to confuse collinear and soft singularities.}. 
Specifically, the terms $|M_1|^2$
and $|M_3|^2$ are singular while the rest are finite. 
These two, unlike for example $|M_4|^2$, involve only two-point
functions of Wilson lines. To 
leading logarithmic accuracy one can
ignore the diagrams giving finite terms and keep only the
contributions of diagrams which are singular, given by 
\be
|M_1|^2 &=& 16 \, (p^-)^2 \,{ z(1+z^2) \over [(1-z) q_t - z k_t]^2} \,
\int d^2r_t\; e^{i (q_t + k_t)\cdot r_t}\, H_F (r_t) \nonumber \\
|M_3|^2 &=& 16 \, (p^-)^2 \,{ z(1+z^2) \over q_t^2} \, \int d^2r_t \;
e^{i (q_t + k_t)\cdot r_t}\, H_A (r_t)
\label{eq:sm1m3}
\ee
with 
\be
H_F (r_t) &\equiv& C_F \, \int d^2b \, {\rm Tr_c}\; 
       \langle [V^{\dagger}(b - r_t/2) -1] \,
               [V(b + r_t/2) -1]\rangle~, \nonumber \\
H_A (r_t) &\equiv& {1\over 2}  \int d^2b \, {\rm Tr_c}\; 
       \langle[U^{\dagger}(b  - r_t/2) -1] \, 
              [U(b + r_t/2) -1]\rangle~,
\label{eq:H_def}
\ee
where ${\rm Tr_c}$ denotes the trace over color matrices.

Integrating over the quark transverse momentum $q_t$, including the phase space
factors and averaging (summing) over initial (final) state degrees of
freedom leads to
\be
\xi {d\sigma^{qA \rightarrow g X} \over d\xi d^2k_t} = {1\over
  (2\pi)^2} \, \xi P_{g/q}(\xi)
\;  {\alpha_s \over 2\pi}\; \log \frac{Q^2}{\Lambda^2} \; \; \int d^2r_t \, 
e^{i k_t\cdot r_t} \, 
\bigg[ H_F(\xi r_t) + H_A(r_t)\bigg]
\label{eq:cs_H}
\ee
where $Q^2$ denotes the factorization scale. In~(\ref{eq:cs_H})
we have introduced the Leading Order (LO) quark-gluon splitting
function~\cite{esw}
\be
\xi P_{g/q}(\xi) \equiv C_F  \left[{1 + (1-\xi)^2 }\right]~.
\label{eq:LO_P}
\ee

It is easy to understand the origin of the two terms
in~(\ref{eq:cs_H}); the first term from Fig.~1-1 corresponds to QED-like
bremsstrahlung, where a free collinear gluon is emitted after  
the quark scatters (multiply) from the target. This contribution vanishes
in the recoilless approximation, i.e.\ for $z=1-\xi\to1$, and
has been neglected in previous computations of gluon production in the color
glass condensate approach~\cite{CGCg_norecoil1,CGCg_norecoil2,CGCg_norecoil3}.
As we shall show, it is important when $z\sim\xi$,
i.e.\ for production of gluons with large rapidity or Feynman-$x$.

The second term, from Fig.~{1-3}, corresponds to the case when the incoming 
quark radiates a collinear gluon which in turn scatters (multiply) from the
target. This is the only surviving contribution for soft radiation,
$\xi\to0$, inherent in the recoilless approximation.  The effect of
recoil for this term is to simply replace the $\xi\to0$ limit of the LO
splitting function $P_{g/q}$ by its exact form given in
(\ref{eq:LO_P}), as expected. This leads to a suppression of
high-energy (forward) radiation by a factor of 2.

To assess the QED-like
contribution $\sim H_F$ from a different perspective,
note that the invariant mass of the intermediate quark propagator
in that diagram (see Fig.~\ref{fig:diagrams}) is given by
\be
m^2 = k_t^2 {z\over\xi} + q_t^2{\xi\over z} - 2q_t\cdot k_t~.
\ee
This diverges for gluons in the central rapidity region, $\xi\to0$,
and with fixed $k_t$, but this diagram
vanishes anyways in that limit. On the other hand, for $\xi\simeq z$ we
have $m^2\simeq (k_t-q_t)^2$, which is not proportional to energy and
becomes small when $k_t$ is nearly collinear to $q_t$; this kinematic
configuration then gives a contribution that is not suppressed by
powers of energy.
The case $\xi\gg z$, finally, corresponds to ``inverse Compton
scattering'' kinematics: a high-energy quark scatters from one or more
small-$x$ gluons and emits a high-energy gluon which
takes over most of the light-cone momentum of the incident
quark~\cite{invCompton}. In order that $m^2$ not be large
(proportional to energy), the quark has to remain nearly collinear to
the beam: $q_t\lsim \sqrt{z/\xi}\, q^-$. For collinear gluons with
$k_t\sim \xi q_t/z\sim \sqrt{\xi/z}\, q^-$ then, $m^2\simeq0$.

The collinear logarithm in the first term of~(\ref{eq:cs_H})
can be understood as
part of the one loop correction to the fragmentation function of the scattered
quark while the collinear logarithm in the second term corresponds to
part of the 
one loop correction to the gluon distribution function. To proceed further, 
we note that there are
pieces in~(\ref{eq:cs_H}) which contribute only when the produced
gluon has zero transverse momentum.  Since we are interested in
gluons with finite transverse momentum, those pieces can be
discarded by defining fundamental and adjoint dipole cross sections at
impact parameter $b$ as follows
\be 
N_F (r_t,b) &\equiv & {1\over N_c} \, {\rm Tr_c} \,
\langle V^{\dagger}(b - r_t/2) V(b + r_t/2) -1\rangle \nonumber \\ 
N_A(r_t,b) &\equiv & {1\over N^2_c -1} \, {\rm Tr_c} \;
\langle U^{\dagger}(b -r_t/2) U(b + r_t/2) -1\rangle
\label{eq:N_FA}
\ee
in terms of which the cross section~(\ref{eq:cs_H}) can be rewritten as
\be
\xi {d\sigma^{qA \rightarrow g X} \over d\xi \,d^2k_t\,d^2b} =
{1 \over  (2\pi)^2} \, 
\xi P_{g/q} (\xi) \, {\alpha_s \over 2\pi}\, \log \frac{Q^2}{\Lambda^2} \;
\bigg[ {1\over \xi^2} N_F(k_t/\xi , b) + N_A(k_t, b)\bigg]~.
\label{eq:cs_N}
\ee

So far, we have calculated the gluon production cross section from
scattering of quarks on the target nucleus. Before we can relate this
to hadron production in proton-nucleus collisions, we need to include 
all other processes, to the same order in $\alpha_s$,
which contribute to hadron production. These are, for example, elastic and
inelastic production of quarks as well as production of gluons from
scattering of gluons on the target nucleus\footnote{The diagrams considered 
here correspond to real corrections, one needs to include the virtual 
corrections to get the full splitting function.}. Details
will be reported elsewhere~\cite{aaj}.  Nevertheless, the final
result is quite simple and can be written down in analogy to
eq.~(\ref{eq:cs_N}):
\be
\xi {d\sigma^{qA \rightarrow g X} \over d\xi \,d^2k_t\,d^2b} &=&
{1 \over  (2\pi)^2} \, 
\xi P_{g/q} (\xi) \, {\alpha_s \over 2\pi}\,\log \frac{Q^2}{\Lambda^2} \;
\bigg[N_A(k_t , b) +  {1\over \xi^2} N_F(k_t/\xi , b)\bigg]
\label{eq:cs_qtog} \\
\xi {d\sigma^{qA \rightarrow q X} \over d\xi \,d^2k_t\,d^2b} &=&
{1 \over  (2\pi)^2} \, 
\xi\, P_{q/q} (\xi) \, {\alpha_s \over 2\pi}\,\log \frac{Q^2}{\Lambda^2} \;
\bigg[N_F(k_t , b) +  {1\over \xi^2} N_F(k_t/\xi , b)\bigg]
\label{eq:cs_qtoq} \\
\xi {d\sigma^{gA \rightarrow q X} \over d\xi \,d^2k_t\,d^2b} &=&
{1 \over  (2\pi)^2} \, 
\xi\, P_{q/g} (\xi) \, {\alpha_s \over 2\pi}\,\log \frac{Q^2}{\Lambda^2} \;
\bigg[N_F(k_t , b) + {1 \over \xi^2} N_A(k_t/\xi , b)\bigg]
\label{eq:cs_gtoq} \\
\xi {d\sigma^{gA \rightarrow g X} \over d\xi \,d^2k_t\,d^2b} &=&
{1 \over  (2\pi)^2} \, 
\xi P_{g/g} (\xi) \, {\alpha_s \over 2\pi}\, \log \frac{Q^2}{\Lambda^2} \;
\bigg[N_A(k_t , b) +  {1\over \xi^2} N_A(k_t/\xi , b)\bigg]
\label{eq:cs_gtog}
\ee
with the LO splitting functions as given in~\cite{esw}. Contributions of
processes involving anti-quarks are identical to those from quarks to
this order. Here, we denote the momentum
fraction of the produced daughter parton by $\xi$, and its
transverse momentum by $k_t$.  For
completeness, we also need to include the elastic scattering
contributions to quark and gluon production~\cite{adjjm_prl} given by
\be
\xi {d\sigma^{q A \rightarrow q X} \over d\xi\,d^2k_t\,d^2b} &=& {2 \over
  (2\pi)^2} \,  \xi\,
\delta(1 - \xi) \, N_F(k_t,b) \label{eq:elastic_qq} \\
\xi {d\sigma^{g A \rightarrow g X} \over d\xi\,d^2k_t\,d^2b} &=& {2 \over
  (2\pi)^2} \, \xi\,
\delta(1 - \xi) \, N_A(k_t,b)
\label{eq:elastic_gg}
\ee

We now reorganize the different pieces from eqs.~(\ref{eq:cs_qtog} -
\ref{eq:elastic_gg}) in a way
which makes the connection to the DGLAP evolution of parton distribution 
and fragmentation functions obvious. We consider the diagrams corresponding 
to the DGLAP evolution of quark distribution functions first. 
They are given by (half of) eq. (\ref{eq:elastic_qq}) and 
the first terms of (\ref{eq:cs_qtoq}) and (\ref{eq:cs_gtoq}).
Multiplying by the bare (parton model) quark and gluon distribution
functions $q_0(x/\xi)$, $g_0(x/\xi)$ we obtain
\begin{eqnarray}
& & \int\limits_x^1 {d\xi\over \xi}
\left\{
q_0({x\over\xi}) 
\left( \delta \left(1 - \xi\right) + {\alpha_s \over 2 \pi} \, \log  
\frac{Q^2}{\Lambda^2} \, 
P_{q/q}\left(\xi\right)\right)
 + 
g_0({x\over\xi}) {\alpha_s \over 2 \pi} \, \log  
\frac{Q^2}{\Lambda^2} \,
P_{q/g}\left(\xi\right)\right\} N_F (\xi,k_t,b) \nonumber\\
& & \to
 f_{q/p} (x/\xi,Q^2) \otimes N_F (\xi,k_t,b)
\label{eq:q_ren}
\end{eqnarray}
where $f_{q/p} (x,Q^2)$ is the renormalized (DGLAP evolved) quark distribution
function in a proton. We have used the DGLAP evolution equation for the quark
distribution function.

Next, consider evolution of the gluon distribution 
function. The relevant terms are given by (half of) eq.~(\ref{eq:elastic_gg}), 
and the first terms of (\ref{eq:cs_qtog}) and (\ref{eq:cs_gtog}). Putting them 
together we obtain
\begin{eqnarray}
& & \int\limits_x^1 {d\xi\over \xi} \left\{ g_0({x\over\xi}) \left[ \delta
\left(1 - \xi\right)
 + {\alpha_s \over 2 \pi} \, \log \frac{Q^2}{\Lambda^2}
\, P_{g/g}\left(\xi\right)\right]
 +q_0({x\over\xi})
{\alpha_s \over 2 \pi} \, \log \frac{Q^2}{\Lambda^2}\,
P_{g/q}\left(\xi\right)\right\}
 N_A (\xi,k_t,b) \nonumber\\
& & \to f_{g/p} (x/\xi,Q^2) \otimes N_A (\xi,k_t,b)
\label{eq:g_ren}
\end{eqnarray}
where $f_{g/p} (x,Q^2)$ is the renormalized (DGLAP evolved) gluon
distribution function in a proton.

The remaining diagrams lead to DGLAP evolution
of the fragmentation functions. We start with the
quark fragmentation function given by  (half of)
eq.~(\ref{eq:elastic_qq}) and the second terms in
eqs.~(\ref{eq:cs_qtog}-\ref{eq:cs_qtoq})
\begin{eqnarray}
& & 
\int\limits_x^1 {d\xi\over \xi} \left[ D^0_q({x\over \xi}) 
\left( \delta (1-\xi) + {\alpha_s \over 2 \pi} \, \log
  \frac{Q^2}{\Lambda^2} \,
 P_{q/q}(\xi)\right) +
D^0_g({x\over \xi}) {\alpha_s \over 2 \pi} \, \log \frac{Q^2}{\Lambda^2} \,
 P_{g/q}(\xi)\right]
 \widetilde{N}_F \left(\xi,k_t,b\right)\nonumber\\
& & \to
D_{q} (x/\xi,Q^2) \otimes \widetilde{N}_F (\xi,k_t,b)
\label{eq:Dq_ren}
\end{eqnarray}
where $D_{q}(x,Q^2)$ is the DGLAP evolved quark-hadron fragmentation
function and $\widetilde{N}_F (\xi,k_t,b)\equiv N_F (\xi,k_t/\xi,b)/\xi^2$.
The rest of the diagrams give 
\begin{eqnarray}
& & \int\limits_x^1 {d\xi\over \xi} \left[ D^0_g({x\over \xi}) 
 \left( \delta (1-\xi) + {\alpha_s \over 2 \pi} \, \log\frac{Q^2}{\Lambda^2} \,
 P_{g/g}(\xi)\right) +
D^0_q({x\over \xi}) {\alpha_s \over 2 \pi} \, \log \frac{Q^2}{\Lambda^2} \,
 P_{q/g}(\xi)\right]
 \widetilde{N}_A (\xi,k_t,b)\nonumber\\
& & \to
D_{g} (x/\xi,Q^2) \otimes \widetilde{N}_A (\xi,k_t,b)
\label{eq:Dg_ren}
\end{eqnarray}
where $\widetilde{N}_A (\xi,k_t,b)\equiv N_A (\xi,k_t/\xi,b)/\xi^2$.
The final result for one parton radiation can be schematically summarized
as the following
\begin{eqnarray}
\rightarrow  f_q(Q^2)\otimes N_F\otimes D_0^q + q_0\otimes\widetilde{N}_F
\otimes D_q(Q^2) 
+ f_g(Q^2)\otimes N_A\otimes D_0^g + g_0\otimes\widetilde{N}_A
\otimes D_g(Q^2)
\label{eq:onerad}
\end{eqnarray}
where $f_{q,g}(Q^2)$, $D_{q,g}(Q^2)$ correspond to DGLAP evolved quark and 
gluon distribution and fragmentation functions, respectively;
$f_0$, $D_0$
are the bare distribution and fragmentation functions, and $\otimes$ denotes
a convolution in $x$ (not shown explicitly).

The expression (\ref{eq:onerad}) is, however, not the complete result as
should be clear from the presence of bare (non-DGLAP evolved)
distribution and fragmentation functions. To get the complete one loop
result, one needs to consider one additional parton radiation.
For example, for the diagram shown in Fig.~{1-1}, one needs to allow
for gluon radiation from the initial quark line also,
which contributes to the one-loop DGLAP
evolution of the bare quark distribution function. This contribution
is calculated in appendix~\ref{appendixA}.
The full calculation will be shown elsewhere~\cite{aaj}, here we just quote 
the final result for hadron production in proton-nucleus collisions
\begin{eqnarray}
{d\sigma^{p A \rightarrow h X} \over dY \, d^2 P_t \, d^2 b} &=& {1
  \over (2\pi)^2}
\int_{x_F}^{1} dx \, {x\over x_F} \Bigg\{
f_{q/p} (x,Q^2)~ N_F [{x\over x_F} P_t , b]~ D_{h/q} ({x_F\over
  x}, Q^2) +  \nonumber \\
&&
f_{g/p} (x,Q^2)~ N_A [{x\over x_F} P_t , b] ~ D_{g/h} ({x_F\over x}, Q^2) 
\Bigg\}
\label{eq:final}
\end{eqnarray}
where $Y$ and $P_t$ are the rapidity and transverse momentum of the produced
hadron while $x_F$ denotes its Feynman-$x$.
Eq.~(\ref{eq:final}) is our main result. It 
should be noted that all additional parton
radiations are taken into account by using the solution of the DGLAP evolution
equations for the quark and gluon distribution and fragmentation functions.    

\section{Forward hadron production in dA collisions at RHIC}

We now apply our results to
deuteron-gold collisions at RHIC. We use the Leading Order CTEQ5
distribution functions for a proton~\cite{cteq} with $Q^2=P_t^2$,
employing isospin symmetry to
obtain those of the neutron, and the LO KKP fragmentation
functions~\cite{KKP} into charged hadrons (divided by two). For a
deuteron projectile, the difference to fragmentation into negatively
charged hadrons is small and nearly independent of transverse momentum.

For our numerical results shown below, we employ the following
parameterization\footnote{Our convention for the sign is
  opposite to that of ref.~\cite{dima}, which is due to the
  definitions~(\ref{eq:N_FA}).}
for the dipole cross section from ref.~\cite{dima}:
\be \label{NA_param}
N_A(r_t,y) = \exp\left[ - \frac{1}{4} [r_t^2
  Q_s^2(y)]^{\gamma(y,k_t)}\right] -1~.
\ee
Here,
\be \label{Qs_y}
Q_s(y) = Q_0 \exp [\lambda (y-y_0)/2]~,
\ee
is the saturation momentum of the nucleus at the rapidity $y$ of the
produced parton, which can be obtained from its logitudinal momentum
$x$ and its transverse momentum $k_t=xP_t/x_F$.
The reference point $y_0=0.6$ specifies where effects due to
quantum evolution start to become important. $Q_0$ is
the initial condition for the saturation momentum at $y_0$, and the growth
rate $\lambda\approx0.3$.

Since data for central collisions is not
publicly available yet, we focus on minimum-bias collisions.
We take an average saturation momentum near midrapidity of
$Q_0=1$~GeV~\cite{dima}. The scattering amplitude for a dipole in the
fundamental representation, $N_F$, can also be parameterized as
in~(\ref{NA_param}), with the replacement $Q_s^2\to Q_s^2\, C_F/C_A =
\frac{4}{9} Q_s^2$.

\begin{figure}[htb]
\centering
\centerline{\epsfig{figure=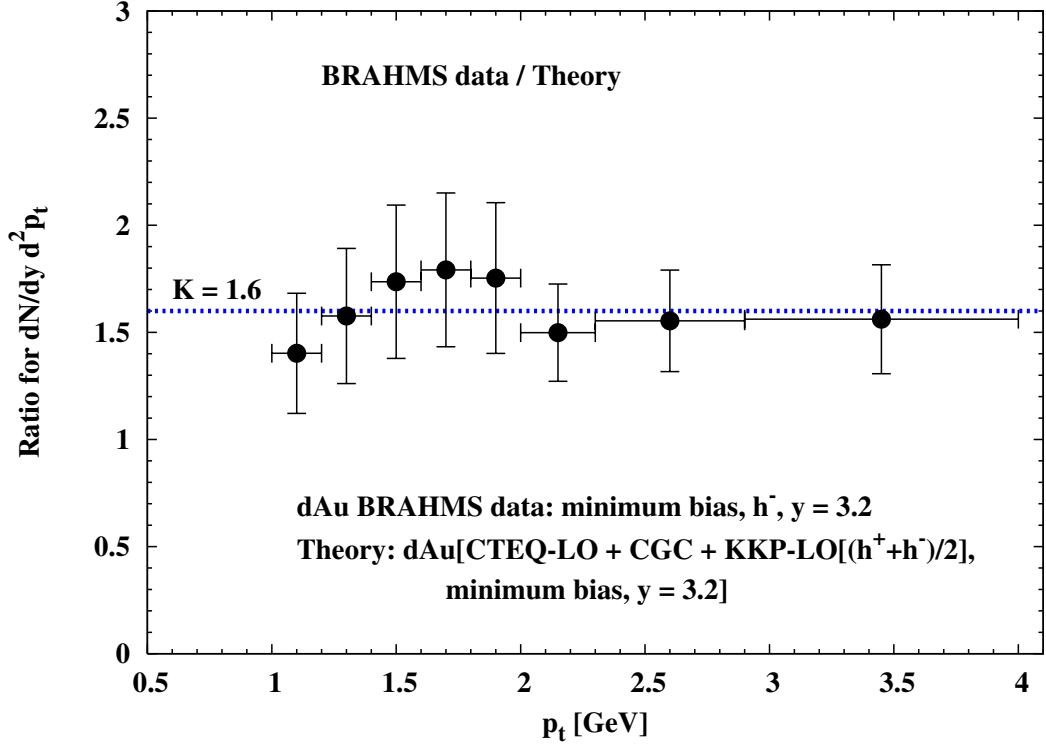,width=4in,angle=-90}}
\caption{$P_t$ dependence of our results compared to BRAHMS minimum
  bias data.}
\label{fig:fig1}
\end{figure}
In~(\ref{NA_param}), $\gamma(y,k_t)$ denotes the anomalous dimension
with saturation boundary condition. Ref.~\cite{dima} employed the
following model for
 the anomalous dimension which reduces to the DGLAP anomalous
dimension (=1) in the high transverse momentum limit:
\be
\gamma(y,k_t) = \frac{1}{2}\left(1+
\frac{|\xi(y,k_t)|}{|\xi(y,k_t)|+\sqrt{2|\xi(y,k_t)|}+28\zeta(3)}
   \right)~,
\ee
where
\be
\xi(y,k_t) = \frac{\log (k_t^2/Q_0^2)}{(\lambda/2)(y-y_0)}~.
\ee
This function vanishes for fixed $k_t$ and
$y\to\infty$, and so $\gamma\to1/2$.
On the other hand, near the boundary to the quantum evolution regime
(i.e.\ at small rapidity $y\simeq y_0$ such that $Q_s\simeq Q_0$) the
anomalous dimension $\gamma\to1$, as appropriate for the classical
McLerran-Venugopalan (MV) model~\cite{MV}. 
For both $\gamma=1/2$ and $\gamma=1$ the Fourier transform
of $N_A(r_t)$ can be performed analytically:
\be
N_A^{\gamma=1}(k_t^2) 
    &=& \int d^2r_t\, e^{i k_t\cdot r_t} N_A^{\gamma=1}(r_t) =
\frac{4\pi}{Q_s^2} \, \exp{(-k_t^2/Q_s^2)}~,\\
 N_A^{\gamma=1/2}(k_t^2) &=&
\int d^2r_t\, e^{i k_t\cdot r_t} N_A^{\gamma=1/2}(r_t) =
\frac{32\pi}{Q_s^2} \frac{1}{(1+16 \, k_t^2/Q_s^2)^{3/2}}~. \label{power}
\ee
Hence, there is an exponential drop with $k_t^2$
for the MV model~\cite{expMV} with
$\gamma=1$, and a power-law decrease if the anomalous dimension
$\gamma=1/2$. Therefore, we expect a steeper $P_t$-distribution of
hadrons for the classical MV model.
In the general case, when $\gamma$ depends on $k_t$ and $y$,
the Fourier transform 
has to be done numerically.

In Fig.~\ref{fig:fig1}, we show the transverse momentum dependence
of the $h^-$ data by BRAHMS~\cite{Arsene:2004ux} relative to our
results, on a linear scale.
The transverse momentum dependence of the data is reproduced very well. For
the correct absolute normalization, we need to multiply by a
$K$-factor of about 1.6; this is not surprising since our calculation
has been performed at Leading Order (in $\alpha_s$).  In fact, from
our point of view it is rather comforting that there is room for the
expected large NLO corrections.

\begin{figure}[htb]
\centering
\centerline{\epsfig{figure=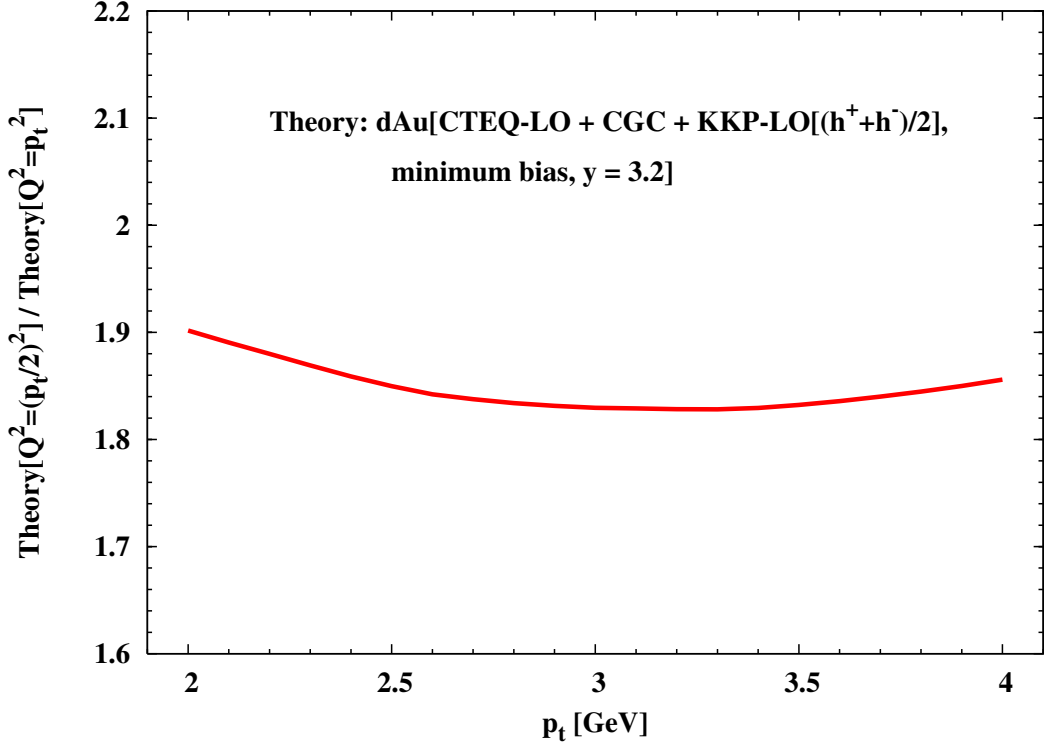,width=4in,angle=-90}}
\caption{Ratio of the transverse momentum distributions of charged
  hadrons for $Q^2=(P_t/2)^2$ and $Q^2=P_t^2$.}
\label{fig:fig_scale}
\end{figure}
To test the scale dependence of our LO result, we plot the ratio of
the distributions for $Q^2/P_t^2=1$ and $Q^2/P_t^2=1/4$ in
Fig.~\ref{fig:fig_scale}. For LO computations there is, in general, a
monotonic dependence on the hard scale and so ``optimal
scales''~\cite{aurenche}, where the result is stable against small
variations of $Q^2/P_t^2$, can not be defined. Nevertheless, we
observe that for $Q^2/P_t^2=1/4$~\cite{aurenche} the shape of the
distribution does turn out to be nearly the same as for $Q^2/P_t^2=1$,
and that the overall $K$-factor drops to $\sim1$. This indicates that NLO
corrections should not change the shape of the distribution by much.
In what follows, we return to the generic scale $Q^2=P_t^2$ and fix $K=1.6$.

\begin{figure}[htb]
\centering
\centerline{\epsfig{figure=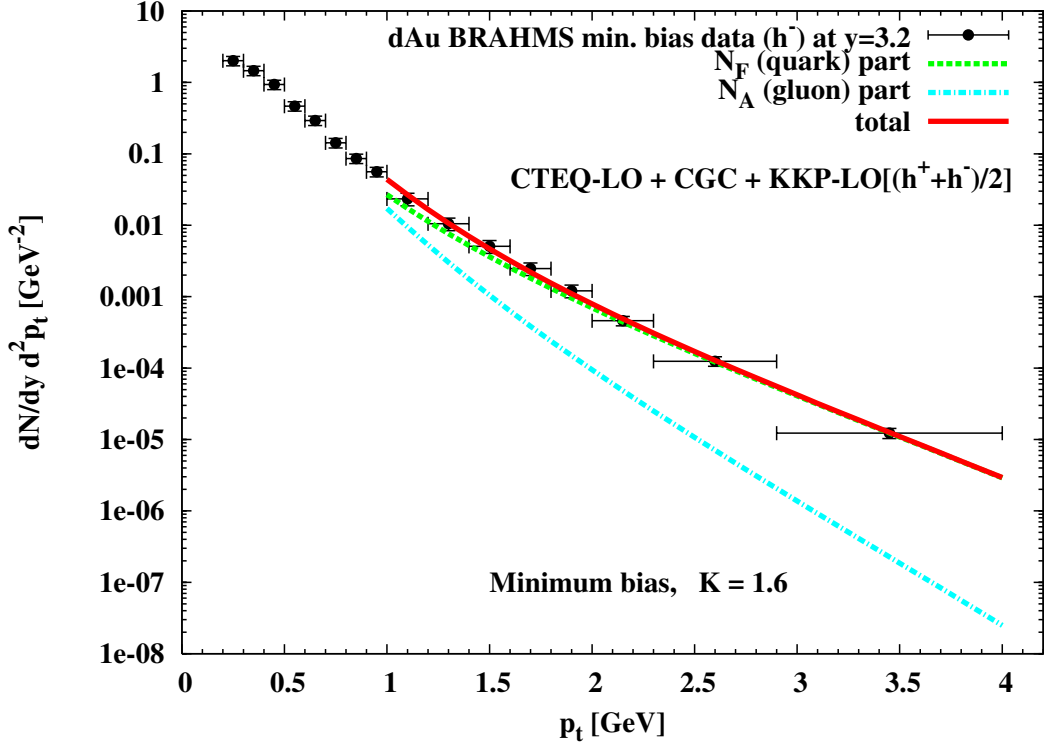,width=4in,angle=-90}}
\caption{Relative contribution of quarks and gluons from the
  projectile deuteron.}
\label{fig:fig2}
\end{figure}
Next, we consider the relative contributions of quarks and gluons
in Fig.~\ref{fig:fig2}. At rapidity $\sim3$ and $p_t\ge1$~GeV
quarks clearly dominate~\cite{adjjm_prl}. 
This, of course, is an essential difference
to the central rapidity region, where gluons contribute more. At yet
larger rapidity quarks would dominate even at lower $p_t$, and their
independent fragmentation should lead to a downward shift of
baryon-number, which is initially concentrated about beam rapidity~\cite{DGS}.

\begin{figure}[hbt]
\centering
\centerline{\epsfig{figure=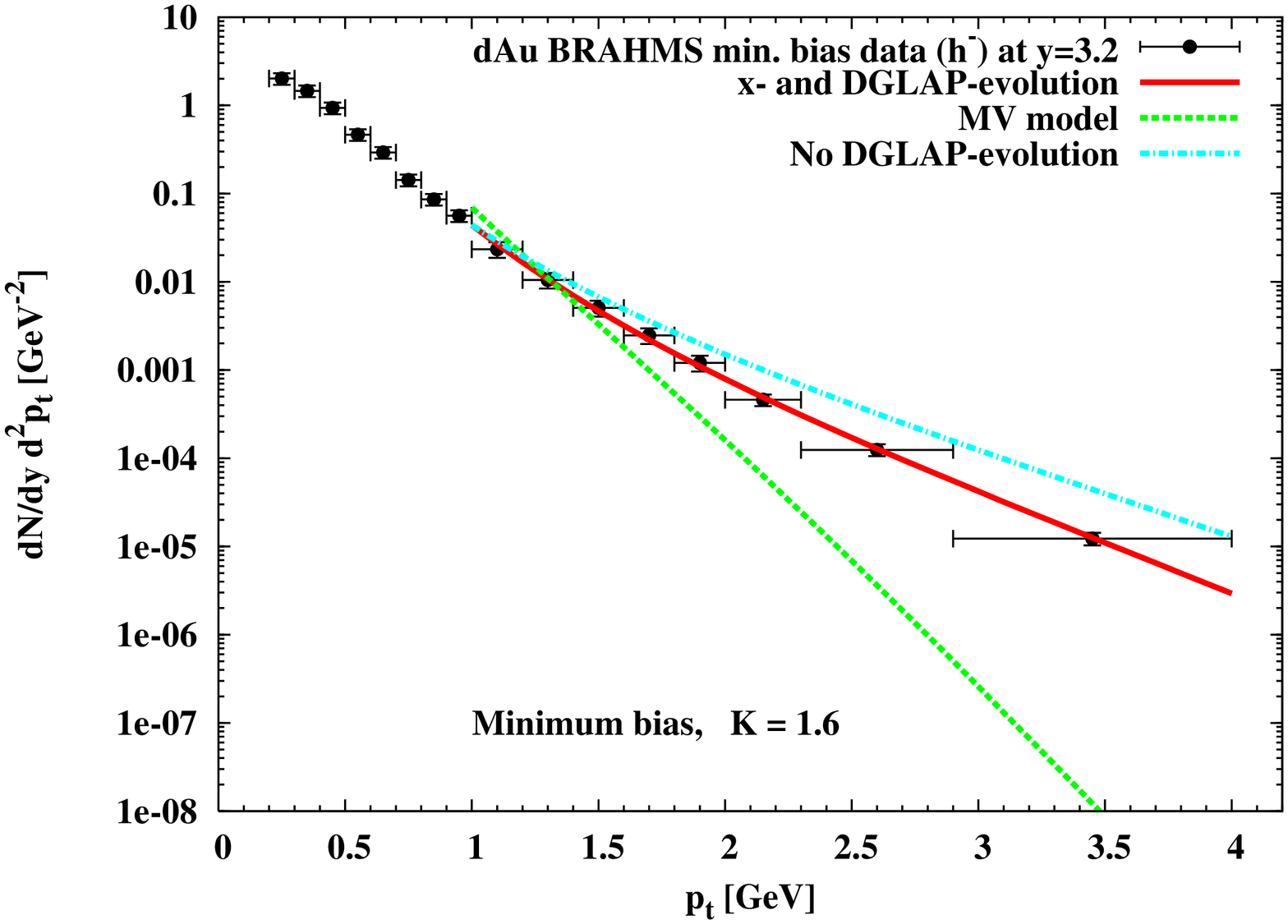,width=4in,angle=-90}}
\caption{Importance of DGLAP evolution of the proton/hadron
  distribution/fragmentation
functions and of the anomalous dimension of the target gluon distribution.}
\label{fig:fig3}
\end{figure}
In Fig.~\ref{fig:fig3}, we compare the full calculation with DGLAP
evolution of the distribution and fragmentation functions to one where
$Q^2$ has been fixed to 1~GeV$^2$. It is obvious that DGLAP evolution is
important and that it improves the agreement with the data
significantly. Physically, this is because collinear parton radiation
shifts the hard partons to smaller
momentum fractions, which is the above-mentioned recoil effect, and
softens the $p_t$-distribution of produced hadrons. The effect is
clearly seen in the data. This emphasizes that full splitting
functions, rather than their soft recoilless limit, have to be employed for
hadron production in the forward region.

\begin{figure}[hbt]
\centering
\centerline{\epsfig{figure=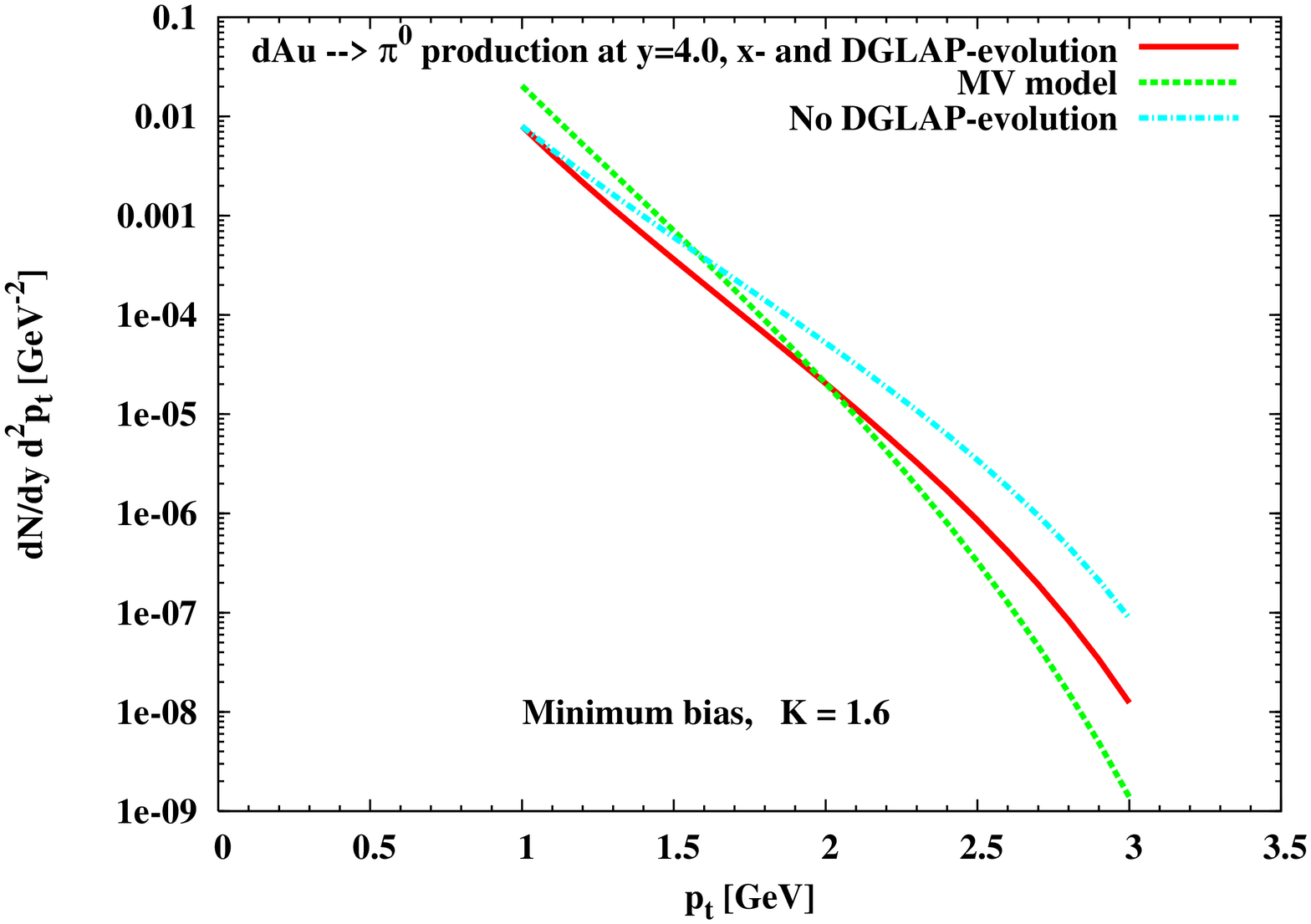,width=4in,angle=-90}}
\caption{Same as Fig.~\protect\ref{fig:fig3} but for $\pi^0$ production at
  rapidity $y=4$.}
\label{fig:fig4}
\end{figure}
We also show the result of a calculation within the classical
McLerran-Venugopalan model, which assumes that the anomalous dimension
$\gamma=1$. The resulting distribution of 
hadrons is much too soft. It is clear that the data requires the proper 
quantum evolution of the target density, with an anomalous dimension
$\gamma$ close to 1/2 (with only a weak dependence on transverse
momentum). This feature is shared by the KKT~\cite{dima} and
IIM~\cite{IIM} dipole models, both of which provide a good description
of the BRAHMS data. 

Fig.~\ref{fig:fig4} shows the $\pi^0$ distribution in minimum-bias
$d+Au$ collisions at rapidity $y=4$, which is currently investigated
by STAR. Here, $x$-evolution leads to an increase of the saturation
momentum by $\sim13\%$ as compared to the BRAHMS kinematics. The
projectile partons need to carry yet larger momentum fractions (and
fragment into faster hadrons), hence $Q^2$-evolution due to emissions
has an even stronger ``softening'' effect on the final hadronic
spectra (roughly one order of magnitude at $p_t=3$~GeV).  The
$P_t$-distribution at $y=4$ is close to exponential for
$P_t:~1\to2.5$~GeV. At higher transverse momentum the convex behavior
of the projectile parton distribution and fragmentation functions
eventually takes over. However, due to the very steep distribution the
relative deviation of the classical MV model from the full result
including quantum evolution in $x$ is smaller than for the BRAHMS
kinematics. Forthcoming STAR data will test this prediction (for
preliminary data at intermediate $P_t$ see~\cite{GregRakness}).

\section{Summary}
In summary, we have shown that high-energy proton-nucleus collisions
can be described as scattering of collinearly factorized
partons (which evolve according to the DGLAP equation) in the
proton on a dense nucleus treated as a Color Glass Condensate.
We have isolated diagrams with collinear singularities which lead to
logarithms of $Q^2$ and proven that they satisfy DGLAP evolution. Only
two-point functions of Wilson lines, i.e.\ dipoles, contribute
to the single-inclusive cross-section;
in the future, these could be obtained from the
JIMWLK small-$x$ evolution equation.

To apply our results to data from RHIC, we have presently adopted a
phenomenological dipole parametrization due to Kharzeev, Kovchegov and
Tuchin. The minimum-bias data from $d+Au$ collisions at forward
rapidity obtained by the BRAHMS collaboration can be reproduced very
well with a transverse momentum {\em independent} $K$-factor of 1.6
(for $Q^2=P_t^2$). Hence, NLO corrections are expected to be large but
should not distort the shape of the transverse momentum distributions
obtained from our LO analysis.

We have shown that in order to reproduce the shape of the measured
transverse momentum distribution, one needs to account for
both $Q^2$-evolution of the projectile parton distribution functions
and of the fragmentation functions, as well as for quantum evolution of
the small-$x$ field of the target nucleus. Neglect of ``recoil''
effects from DGLAP evolution leads to a significantly harder $p_t$
distribution than the data; a purely classical target field with
anomalous dimension $\gamma=1$ produces a much softer
distribution. The fact that both of these key components, and of their
interplay, is seen in the data represents important
evidence for the Color Glass Condensate from RHIC.

In the future, it would be useful if {\em central} rather than
minimum-bias data were available. For central collisions the
saturation momentum of the nucleus increases by $\sim50\%$
and so the saturation
regime extends to higher transverse momentum. Also, it will be
interesting to see whether forthcoming data from STAR for $\pi^0$
production at yet larger rapidity, $y\sim4$, can be reproduced equally
well; we have provided a prediction based on the current setup. A
better $p_t$-resolution will be important to constrain the dipole
profiles more tightly. Quantitative theoretical ab-initio computations of those
universal CGC functions will hopefully emerge in the near future.
We have provided a systematical framework which will make it possible
to compare them to experimental data.

\vspace{0.3in}
\section*{Acknowledgments}
%\leftline{\bf Acknowledgments} 

We would like to thank L.~Frankfurt, F.\ Gelis, M.~Strikman and W.\
Vogelsang for useful discussions.
J.J-M.\ is supported in part by the U.S.\ Department of Energy under 
Grant No.\ DE-FG02-00ER41132.

\begin{appendix}

\section{Radiation from both initial and final state}  \label{appendixA}

\begin{figure}[htp]
\centering
\setlength{\epsfxsize=6cm}
\centerline{\epsffile{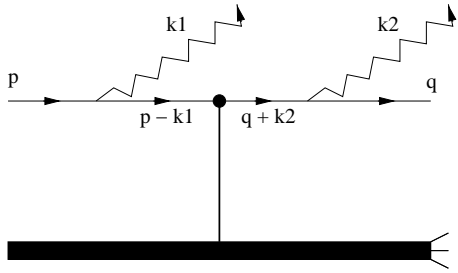}}
\caption{One of the diagrams contributing to DGLAP evolution of bare quark
distribution and fragmentation functions.}
\label{fig:tworad-lab}
\end{figure}
The diagram considered here is shown in Fig.~\ref{fig:tworad-lab}.
The cross section for this process is
\begin{eqnarray}
q^- {d\sigma^{qA \rightarrow q X} \over d q^- d^2 q_t d^2 b} = {1\over 2 p^-}
{1\over 2 (2\pi)^3} \int {d^3 k_1 \over (2\pi)^3 2 k_1^-}   
\int {d^3 k_2 \over (2\pi)^3 2 k_2^-} \, |M|^2
\label{eq:cs-twogl}
\end{eqnarray}
which can be related to hadron production cross section via
\begin{eqnarray}
P^-_h {d\sigma^{pA \rightarrow h X} \over d P^-_h d^2 P_t d^2 b} 
\equiv \int dx_q \, q_0 (x_q) \int dz_h \, D_0 (z_h) \,  
q^- {d\sigma^{qA \rightarrow q X} \over d q^- d^2 q_t d^2 b}~.
\label{eq:cs-conv}
\end{eqnarray}
We have defined $x_q = p^-/P^-, z_h = P^-_h/q^- $ and $p^-, P^-$
are the momenta of the incoming quark and proton respectively, while 
$q^-, P^-_h$ are the momenta of the outgoing quark and hadron. We also
define $z_1= (p^- -k_1^-)/p^-$ and $z_2 = q^-/(q^- + k_2^-)$. The matrix 
element is given by
\begin{eqnarray}
M^{\mu\nu} = g^2 \bar{u} (q) \gamma^{\nu} t^b S_F^0 (q+k_2) 
\tau_f (q + k_2, p - k_1 ) S_F^0 (p-k_1) \gamma^{\mu} t^a u (p)~,
\label{eq:M_munu}
\end{eqnarray}
where
\begin{eqnarray}
\tau_f (q,p)\equiv (2\pi)\delta(p^- - q^-) \,\gamma^-\, \int d^2x\, 
e^{i (q_t - p_t)\cdot x_t}\, [V(x_t) - 1]~.
\label{eq:tau_f}
\end{eqnarray}

We need to contract this with the polarization vectors for the two radiated
gluons $\epsilon_{\mu} (k_1,\lambda_1), \epsilon_{\nu}(k_2,\lambda_2)$ and 
then square it. Using the polarization tensor 
$\sum_{\lambda} \epsilon_{\mu}^{\star}(k,\lambda) \epsilon_{\nu}(k,\lambda) 
\equiv [-g_{\mu\nu} + {\eta_{\mu} k_{\nu} + \eta_{\nu} k_{\mu} \over \eta \cdot k}]$,
we get 
\begin{eqnarray}
|M|^2 &=& {g^4 \over 2} {1\over (q+k_2)^4} {1\over (p-k_1)^4} 
[-g_{\mu\delta} + {\eta_{\mu} k_{1\delta} + \eta_{\delta} k_{1\mu} 
\over \eta \cdot k_1}]
[-g_{\nu\rho} + {\eta_{\nu} k_{2\rho} + \eta_{\rho} k_{2\nu} 
\over \eta \cdot k_2}]
\nonumber\\
&&Tr_D \bigg[
{\slq} \gamma^{\nu} ({\slq} + {\slk}_2) \gamma^- ({\slp} - {\slk}_1) \gamma^{\mu}
{\slp} \gamma^{\delta} ({\slp} - {\slk}_1) \gamma^- ({\slq} + {\slk}_2) \gamma^{\rho}
\bigg]
\nonumber\\
&&
Tr_c \bigg[ t^b t^b [V(q_t + k_{2t} + k_{1t} - p_t) - 
(2\pi)^2 \delta^2 (q_t + k_{2t} + k_{1t} - p_t)] 
\nonumber \\
&&t^a t^a 
 [V^{\dagger}(q_t + k_{2t} + k_{1t} - p_t) - 
(2\pi)^2 \delta^2 (q_t + k_{2t} + k_{1t} - p_t)]\bigg]
\label{eq:Msq}
\end{eqnarray}
where $Tr_D, Tr_c$ stand for traces over spinor and color matrices.
Evaluating the trace of the spinors above is algebraically involved but 
straightforward. Contracted with the gluon polarization tensors, it 
is given by 
\begin{eqnarray}
[-g_{\mu\delta} + \cdots]
[-g_{\nu\rho} + \cdots] \, Tr_D [\cdots] = 
124\, p^- q^- (p\cdot k_1) (q\cdot k_2) 
{ (1+z_1) (1+z_2) \over z_2 (1-z_1) (1-z_2)}
\label{eq:tr_D}
\end{eqnarray}
where $z_1, z_2$ are defined above. Putting everything together, the partonic
cross section can be written as 
\begin{eqnarray}
&&q^- {d\sigma^{q A \rightarrow q X} \over dq^- d^2 q_t d^2 b} = 
{g^4 C_F^2 \over (2\pi)^4} {q^- P^-} \int {dz_1 \over z_1}
{1+ z_1^2 \over 1-z_1} 
\int {dz_2 \over z_2^3} {1+ z_2^2 \over 1-z_2}\, 
\delta (x_q - {q^- \over z_1 z_2 P^-}) 
\int d^2 x_t \, d^2 y_t \nonumber \\ 
&& e^{i q_t \cdot (x_t - y_t)} 
\int {d^2 k_{1t} \over (2\pi)^2} 
{e^{i k_{1t} \cdot (x_t - y_t)} \over k_{1t}^2}
\int {d^2 k_{2t} \over (2\pi)^2}  
{e^{i k_{2t} \cdot (x_t - y_t)} \over [k_{2t} - {1-z_2 \over z_2} q_t]^2}\,
Tr_c \bigg[ [V(x_t) -1 ] [V^{\dagger} (y_t) -1]\bigg]
\label{eq:cs_part}
\end{eqnarray}
Again, the integrations over the transverse momenta $k_{1t},k_{2t}$
exhibit collinear singularities which
lead to logarithms of $Q^2$, so that the cross section can be written as
\begin{eqnarray}
&&q^- {d\sigma^{q A \rightarrow q X} \over dq^- d^2 q_t d^2 b} = 
{1\over (2\pi)^2} {q^-\over P^-} 
\int {dz_1 \over z_1} \left[{\alpha_s\over 2\pi} \ln {Q^2\over \Lambda^2} 
C_F {1+ z_1^2 \over 1-z_1}\right] 
\int {dz_2 \over z_2^3} \left[{\alpha_s\over 2\pi} \ln {Q^2\over \Lambda^2} 
C_F {1+ z_2^2 \over 1-z_2}\right] \nonumber \\
&&\delta (x_q - {q^- \over z_1 z_2 P^-}) \, N_F [{q_t\over z_2}, b]
\label{eq:cs_part2}
\end{eqnarray}
Using (\ref{eq:cs_part2}) in (\ref{eq:cs-conv}) and combining it with the 
relevant term in (\ref{eq:onerad}) gives the quark contribution to the 
hadronic cross section
\begin{eqnarray}
&&{d\sigma^{pA \rightarrow h X} \over dy d^2 P_t d^2 b}
= x_F \int {dz_1 \over z_1} {dz_2\over z_2} {dz_h \over z_h} \, 
q_0 [{x_F \over z_1 z_2 z_h}] 
\left[\delta (1-z_1) + 
{\alpha_s\over 2\pi} \ln {Q^2\over \Lambda^2} P_{q/q}(z_1)\right]\nonumber \\
&&D_0(z_h) 
\left[\delta (1-z_2) + 
{\alpha_s\over 2\pi} \ln {Q^2\over \Lambda^2} P_{q/q}(z_2)\right]
{1\over (2\pi)^2} {1 \over (z_2 z_h)^2} N_F [{P_t \over z_2 z_h}, b] 
\label{eq:cs_had2}
\end{eqnarray}
Defining the DGLAP evolved quark distribution function 
$f_q ({x_F\over z_2 z_h}, Q^2)$ as
\begin{eqnarray}
f_q ({x_F\over z_2 z_h}, Q^2) \equiv \int {dz_1 \over z_1} 
q_0 ({x_F\over z_1 z_2 z_h}) \left[\delta (1-z_1) + 
{\alpha_s\over 2\pi} \ln {Q^2\over \Lambda^2} P_{q/q}(z_1)\right]~,
\label{eq:f_qevo}
\end{eqnarray}
then changing variables to $z^{\prime}_h \equiv z_2 z_h$, and 
defining the DGLAP evolved fragmentation function 
$D_q (z^{\prime}_h,Q^2)$ as
\begin{eqnarray}
D_q (z^{\prime}_h, Q^2) \equiv \int {dz_2 \over z_2} 
D_0 ({z_h^{\prime}\over z_2}) \left[\delta (1-z_2) + 
{\alpha_s\over 2\pi} \ln {Q^2\over \Lambda^2} P_{q/q}(z_2)\right]
\label{eq:D_qevo}
\end{eqnarray}
leads to the DGLAP evolved hadron production cross section
\begin{eqnarray}
&&{d\sigma^{pA \rightarrow h X} \over dy d^2 P_t d^2 b}
= \int_{x_F}^{1} dz_h^{\prime} \, {x_F \over z_h^{\prime}} \,
f_q ({x_F \over z_h^{\prime}}, Q^2) \, D_q (z_h^{\prime}, Q^2)
{1\over (2\pi)^2} {1\over z_h^{\prime 2}} N_F [{P_t\over z_h^{\prime}}, b]~.
\label{eq:cs_hadfin}
\end{eqnarray}
It should be noted that one also needs to include diagrams where 
one integrates over the final state quark momenta rather than the 
gluon momenta as we have done here. This would bring in the quark-gluon
splitting function $P_{g/q}$ which, combined with $P_{q/q}$ above, would 
then go into the DGLAP evolution of the quark distribution or fragmentation 
function in~(\ref{eq:f_qevo}, \ref{eq:D_qevo}). A further change of variables
$x = {x_F / z_h^{\prime}}$ in~(\ref{eq:cs_hadfin}) then gives the 
first part of~(\ref{eq:final}). Contribution of the diagrams involving an 
incoming gluon are similar and give the second part of~(\ref{eq:final}).

\section{$2\to1$ kinematics}  \label{appendixB}

\begin{figure}[htp]
\centering
\setlength{\epsfxsize=6cm}
\centerline{\epsfig{figure=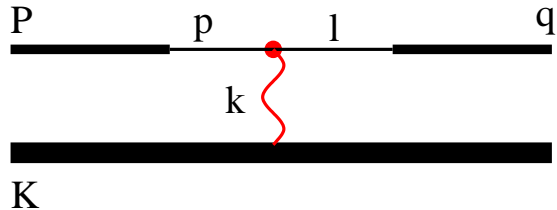,width=3in}}
\caption{The $2 \rightarrow 1$ kinematics.}
\label{fig2to1}
\end{figure}

In this appendix we elaborate on the $2\to1$ like kinematics employed
above. In particular, we show that energy-momentum conservation
implies that much smaller momentum fractions are probed in the target
than for $2\to2$ kinematics underlying leading-twist perturbative
computations (see e.g.~\cite{WernerMark}). We work in a frame where
both projectile and target have large light-cone momenta, for example
the center of rapidity frame. The various momenta as used within this
appendix are defined in Fig.~\ref{fig2to1}.
The incoming parton carries momentum $p$ while that of the 
projectile nucleons is labeled $P$; the outgoing parton's momentum is $l$
while that of the outgoing hadron is $q$. The
momentum of the nucleons from the incoming nucleus is $K$, and $k$ is
the exchange between the parton and the nucleus. More
explicitly\footnote{We follow the convention from the main text that
  $P$ and $q$ have large minus-components, i.e.\ that the rapidities
  of both the projectile and of the produced hadron are large and
  negative. To comply with standard practice, however, in the final
  figures we quote the modulus of the rapidity of the detected hadron,
$y_h\equiv-Y>0$.},
\begin{eqnarray}
P^{\mu} &=& (P^-=\sqrt{s/2}, P^+=0, P_t=0)\nonumber \\
p^{\mu} &=& (p^- = x \, P^-, p^+=0, p_t=0)\nonumber \\
l^{\mu} &=& (l^- = p^-, l^+ = {l_t^2 \over 2 l^-}, l_t) \nonumber \\
q^{\mu} &=& (q^- = z \, l^-, q^+ = {q_t^2 \over 2 q^-}, q_t = z \,
l_t) \nonumber\\
k^{\mu} &=& (k^- \approx 0, k^+ = x_A \, K^+, k_t) \nonumber \\
K^{\mu} &=& (K^- =0, K^+ = \sqrt{s/2}, K_t=0)
\label{eq:2to1_label}
\end{eqnarray}
and 
\begin{eqnarray}
l = p + k
\label{eq:mom_cons}
\end{eqnarray}
from energy-momentum conservation. The only approximation made 
in (\ref{eq:2to1_label}) is setting  $k^- \approx 0$ which corresponds to
the eikonal approximation. Using~(\ref{eq:mom_cons}), we have $k_t = l_t$, 
$x P^- = l^-$ and $k^+ = l^+$, so that $z= x_F/x$ and, finally,
\begin{eqnarray}
x_A = {x \, q_t^2 \over x_F^2 \, s}~,
\label{eq:x_A_q_t}
\end{eqnarray}
where Feynman-$x$ is defined as $x_F \equiv {q^- / \sqrt{s/2}}$.
This relation can be rewritten in terms of the rapidity of the
produced hadron, $Y$ (which in the massless limit equals that of its
parent parton):
\begin{eqnarray}
x_A \equiv x\, e^{- 2 y_h}~. 
\label{eq:x_A_y}
\end{eqnarray}
Here, $y_h\equiv-Y>0$ denotes the modulus of the rapidity of the
observed final-state hadron.
In terms of its momentum, its rapidity is given by
$q^- \equiv q_t\, e^{y_h} / \sqrt{2}$.

\begin{figure}[htp]
\centering
\setlength{\epsfxsize=6cm}
\centerline{\epsfig{figure=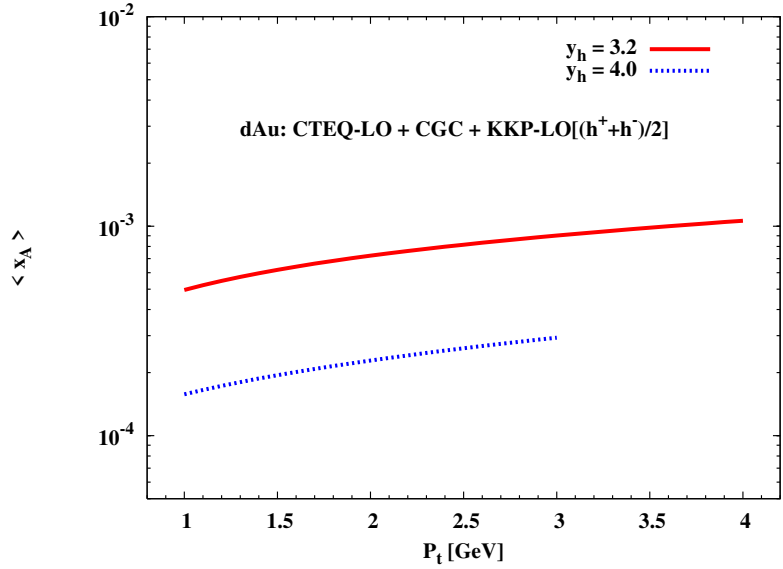,width=3in,angle=-90}}
\caption{The average momentum fraction probed in the target for
  forward hadron production at RHIC energy, versus the
  transverse momentum of the hadron.}
\label{fig_xA}
\end{figure}

Equation~(\ref{eq:x_A_q_t}) relates $x_A$ to the momentum 
fraction $x$ carried by the impinging projectile parton, and to the transverse 
momentum and Feynman-$x$ of the produced hadron. Hence, we can 
insert this form inside the integral from eq.~(\ref{eq:final}), then divide by
eq.~(\ref{eq:final}) itself, to determine the average $x_A$ probed in
the target nucleus. It is clear from~(\ref{eq:x_A_q_t}) that $\langle
        x_A\rangle \to q_t^2/s\sim 10^{-4}$ as $x_F\to 1$ (since $1\ge
        x\ge x_F$).
The result for both BRAHMS and STAR kinematics is shown in Fig.~\ref{fig_xA}.
It turns out that for BRAHMS kinematics for example,
$\langle x_A\rangle\approx
10^{-3}$, which is more than an order of magnitude smaller than for the $2\to2$
kinematics employed in leading-twist calculations~\cite{WernerMark}.

\begin{figure}[htp]
\centering
\setlength{\epsfxsize=6cm}
\centerline{\epsfig{figure=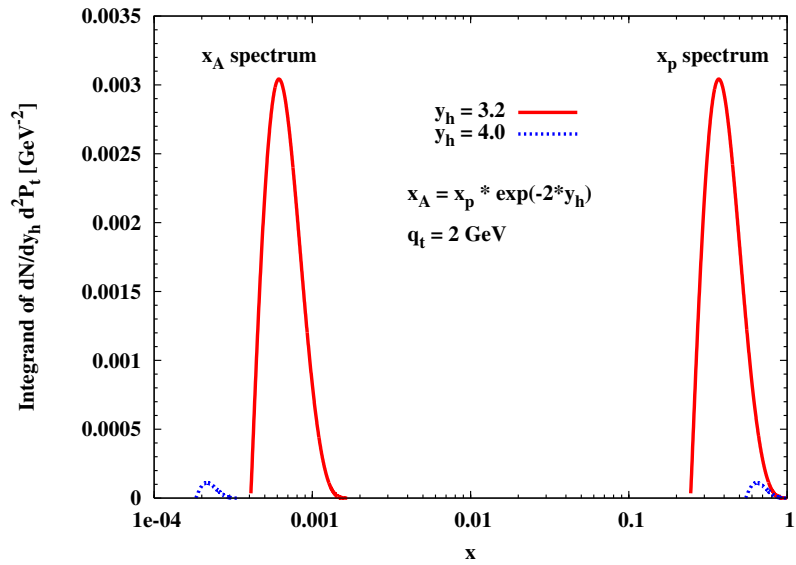,width=3in,angle=-90}}
\caption{The momentum fractions of projectile and target partons
contributing to the cross section
from eq.~(\ref{eq:final}).}
\label{fig:kin}
\end{figure}

In Fig.~\ref{fig:kin} we show the kinematic region in $x$ which 
contributes to the cross section given by (\ref{eq:final}), at 
hadron transverse momentum $q_t=2$~GeV and rapidity $y_h=3.2$ and 4,
respectively. Clearly,
the cross section is dominated by very large $x$ (labeled in this figure
as $x_p$) and very small values of $x_A$.

\end{appendix}
%%%%%%%%%%%%%%%%%%%%%%%%%%%%%%%%%%%%%%%%%%%%%%%%%%%%%%%%%%%%%%%%%%%%%%%%

\end{document}